\documentclass[aps,prd,twocolumn,showpacs,superscriptaddress]{revtex4-1}
\usepackage{amsmath}
\usepackage{latexsym}
\usepackage{amsfonts}
\usepackage{graphicx}
\usepackage{hyperref}
\usepackage{color}

\begin{document}

\title{Simple single field inflation models and the running of spectral index}

\author{Qing  Gao}
\email{gaoqing01good@163.com}
\affiliation{MOE Key Laboratory of Fundamental Quantities Measurement,
School of Physics, Huazhong University of Science and Technology,
Wuhan 430074,  P. R. China }

\author{Yungui Gong}
\email{yggong@mail.hust.edu.cn}
\affiliation{MOE Key Laboratory of Fundamental Quantities Measurement,
School of Physics, Huazhong University of Science and Technology,
Wuhan 430074,  P. R. China }

\author{Tianjun Li}
\affiliation{State Key Laboratory of Theoretical Physics
and Kavli Institute for Theoretical Physics China (KITPC),
Institute of Theoretical Physics, Chinese Academy of Sciences,
Beijing 100190, China}
\affiliation{School of Physical Electronics,
University of Electronic Science and Technology of China,
Chengdu 610054, China}

\author{Ye Tian}

\affiliation{State Key Laboratory of Theoretical Physics
and Kavli Institute for Theoretical Physics China (KITPC),
Institute of Theoretical Physics, Chinese Academy of Sciences,
Beijing 100190, China}

\begin{abstract}

The BICEP2 experiment confirms the existence of primordial gravitational wave with the tensor-to-scalar
ratio $r=0$ ruled out at $7\sigma$ level. The consistency of this large value of $r$ with
the {\em Planck} data requires a large negative running $n'_s$ of the scalar spectral index.
Herein we propose two types of the single field inflation models with simple potentials
to study the possibility of the consistency of the models with the BICEP2 and {\em Planck} observations.
One type of model suggested herein is realized in the supergravity model building.
These models fail to provide the needed $n'_s$ even though both
can fit the tensor-to-scalar ratio and spectral index.

\end{abstract}

\preprint{arXiv: 1404.7214}
\maketitle

\section{Introduction }

The observed temperature fluctuations in the cosmic microwave background radiation (CMB)
strongly suggested that our Universe might experience an accelerated expansion, more precisely,
inflation~\cite{starobinskyfr, guth81, linde83, Albrecht:1982wi}, at a seminal stage of
evolution. In addition to the solution to the problems in the standard big bang cosmology
such as the flatness, horizon, and monopole problems,
the inflation models predict the cosmological perturbations in the matter density
from the inflaton vacuum fluctuations, which describes the primordial
power spectrum consistently. Besides the scalar perturbation, the tensor perturbation is generated as well,
which gives the B-mode polarisation
as a signature of the primordial gravitational wave.

Recently, the BICEP2 experiment has discovered the primordial gravitational wave
with the B-mode power spectrum around $\ell \sim 80$~\cite{Ade:2014xna}. If it is confirmed,
it will seemingly forward the study in fundamental physics.
BICEP2 experiment~\cite{Ade:2014xna} has measured the tensor-to-scalar ratio to be
$r=0.20^{+0.07}_{-0.05}$ at the 68\% confidence level for the lensed-$\Lambda$CDM model, with
$r=0$ disfavoured at $7.0\sigma$ level.
Subtracting the various dust models and re-deriving the $r$
constraint still results in high significance of detection, it results in
$r=0.16^{+0.06}_{-0.05}$. From the first-year observations,
the {\em Planck} temperature power spectrum \cite{planck13} in combination with the nine years of Wilkinson Microwave Anisotropy Probe (WMAP) polarization low-multipole likelihood \cite{wmap9}
and the high-multipole spectra from the Atacama Cosmology Telescope (ACT) \cite{act13} and the South Pole Telescope (SPT) \cite{spt11}
({\em Planck}+WP+highL) constrained the tensor-to-scalar
ratio to be $r \le 0.11$ at the 95\% confidence level \cite{Ade:2013zuv,Ade:2013uln}.
Therefore, the BICEP2 result is in disagreement with the {\em Planck} result.
To reduce the inconsistency between Planck and BICEP2 experiments, we need to
include the running of the spectral index $n'_s=d\ln n_s/d\ln k$. With the running of the spectral index,
the 68\% constraints  from the {\em Planck}+WP+highL data are
$n_s=0.9570\pm 0.0075$ and $n'_s=-0.022\pm 0.010$ with $r<0.26$ at the 95\% confidence level.
Thus, the running of the spectral index needs to be smaller than
0.008 at the 3$\sigma$ level for any viable inflation model.

Because different inflationary models predict different magnitudes for the
tensor perturbations, such large tensor-to-scalar ratio $r$ from the BICEP2 measurement
will give a strong constraint on the inflation models. Also, the inflaton potential is
around the Grand Unified Theory (GUT) scale $2\times 10^{16}$~GeV, and Hubble scale
is about $1.0\times 10^{14}$~GeV. From the naive analysis of Lyth bound~\cite{Lyth:1996im},
a large field inflation will be experienced, and then the validity of effective field theory will
be challenged since the high-dimensional operators are suppressed by the
reduced Planck scale. The inflation models, which can have $n_s\simeq 0.96$ and
$r\simeq 0.16/0.20$, have been studied extensively~\cite{Anchordoqui:2014uua, Czerny:2014wua,
Ferrara:2014ima, Zhu:2014wda, Gong:2014cqa,
Okada:2014lxa, Ellis:2014rxa, Antusch:2014cpa,
Freivogel:2014hca, Bousso:2014jca, Kaloper:2014zba, Choudhury:2014kma,*Choudhury:2014wsa,*Choudhury:2013iaa,*Choudhury:2014kma, Choi:2014aca,
Murayama:2014saa, McDonald:2014oza, Gao:2014fha, Ashoorioon:2014nta,*Ashoorioon:2013eia,*Ashoorioon:2009wa,*Ashoorioon:2011ki,Sloth:2014sga,
Kawai:2014doa,Kobayashi:2014rla,*Kobayashi:2014ooa,*Kobayashi:2014jga,Bastero-Gil:2014oga,DiBari:2014oja,Ho:2014xza,Hotchkiss:2011gz}.
Specifically, the simple chaotic
 and natural inflation models are preferred.

Conversely, supersymmetry is the most promising extension
for the particle physics Standard Model (SM).
Specifically, it can stabilize the scalar masses, and has a non-renormalized
 superpotential. Also, gravity is critical in the
early Universe, so it seems to us that supergravity theory is
a natural framework for inflation model building~\cite{Freedman:1976xh,*Deser:1976eh,Antusch:2009ty,*Antusch:2011ei}. However,
 supersymmetry breaking scalar masses in a generic supergravity theory
are of the same order as the gravitino mass, inducing
the reputed $\eta$ problem~\cite{Copeland:1994vg,*Stewart:1994ts,*adlinde90,*Antusch:2008pn,*Yamaguchi:2011kg,*Martin:2013tda,Lyth:1998xn},
where all the scalar masses are of the order of the Hubble parameter
because of the large vacuum energy density during inflation~\cite{Goncharov:1984qm}.
There are two elegant solutions: no-scale supergravity~\cite{Cremmer:1983bf,*Ellis:1983sf,*Ellis:1983ei,*Ellis:1984bm,*Lahanas:1986uc,
Ellis:1984bf, Enqvist:1985yc, Ellis:2013xoa, Ellis:2013nxa, Li:2013moa, Ellis:2013nka},
and shift-symmetry in the K\"ahler potential~\cite{Kawasaki:2000yn, Yamaguchi:2000vm,
Yamaguchi:2001pw, Kawasaki:2001as, Kallosh:2010ug, Kallosh:2010xz, Nakayama:2013jka,
Nakayama:2013txa, Takahashi:2013cxa, Li:2013nfa}.

Thus, three issues need to be addressed regarding the criteria of the inflation model building: \\

Firstly (C-1), the spectral index is around 0.96, and the tensor-to-scalar ratio is around 0.16/0.20. \\

Secondly (C-2), to reconcile the Planck and BICEP2 results, we need to have $n'_s \sim -0.22$.
For simplicity, we do not consider the alternative approach
here~\cite{Freivogel:2014hca, Bousso:2014jca}. \\

Lastly (C-3), we need to violate the Lyth bound and try to realize the sub-Planckian inflation.
For simplicity, we will not consider the alternative mechanisms such as two-field inflation
models \cite{Hebecker:2013zda,McDonald:2014oza}, and the models which
employ symmetries to control the quantum corrections
like the axion monodromy~\cite{McAllister:2008hb}. \\

It seemingly appears that (C-1) can be satisfied by a considerable amount of inflaton potentials, thus,
this is not a difficulty to overcome. In this paper, we will propose two types
of the simple single field inflation models,
and show that their spectral indices and tensor-to-scalar ratios
are highly consistent with both the Planck and BICEP2 experiments. We construct one type of
 inflation models from the supergravity theory with shift symmetry
in the K\"ahler potential. However, in these simple inflation models,
we will show that (C-2) and (C-3) can not be satisfied.

\section{Slow-roll Inflation}

The slow-roll parameters are defined as
\begin{gather}
\label{slow1}
\epsilon=\frac{M_{pl}^2V_\phi^2}{2V^2},\\
\label{slow2}
\eta=\frac{M_{pl}^2V_{\phi\phi}}{V},\\
\label{slow3}
\xi^2=\frac{M_{pl}^4V_\phi V_{\phi\phi\phi}}{V^2},
\end{gather}
where $M^2_{pl}=(8\pi G)^{-1}$, $V_\phi=dV(\phi)/d\phi$, $V_{\phi\phi}=d^2V(\phi)/d\phi^2$
and $V_{\phi\phi\phi}=d^3V(\phi)/d\phi^3$. For the single field inflation, the scalar power spectrum is
\begin{equation}
\label{power}
\mathcal{P}_{\mathcal{R}}=A_s\left(\frac{k}{k_*}\right)^{n_s-1+n_s'\ln(k/k_*)/2},
\end{equation}
where the subscript "*" means the value at the horizon crossing, the scalar amplitude is thus
\begin{equation}
\label{power1}
A_s\approx \frac{1}{24\pi^2M^4_{pl}}\frac{\Lambda^4}{\epsilon},
\end{equation}
the scalar spectral index and the running are
given \cite{Lyth:1998xn,Stewart:1993bc} by
\begin{gather}
\label{nsdef}
\begin{split}
n_s\approx1+2\eta-6\epsilon+2\left[\frac{1}{3}\eta^2+(8C-1)\epsilon \eta\right.\\
\left.-\left(\frac{5}{3}+12C\right)\epsilon^2-\left(C-\frac{1}{3}\right)\xi^2\right],
\end{split}\\
\label{rundef}
n_s'=16\epsilon\eta-24\epsilon^2-2\xi^2,
\end{gather}
where $C=-2+\ln 2+\gamma \simeq -0.73$ with $\gamma$ the Euler-Mascheroni constant.
The tensor power spectrum is
\begin{equation}
\label{powerr}
\mathcal{P}_{T}=A_T\left(\frac{k}{k_*}\right)^{n_t},
\end{equation}
the tensor spectral index and the tensor to scalar ratio \cite{Lyth:1998xn,Stewart:1993bc} are
\begin{gather}
\label{ntdef}
n_t=-2\epsilon\left[1+\left(4C+\frac{11}{3}\right)\epsilon-2\left(\frac{2}{3}+C\right)\eta\right]\approx -2\epsilon,\\
\label{rdef}
r=\frac{A_T}{A_s}=16 \epsilon \left[ 1+8\left(C+\frac{2}{3}\right)(2\epsilon-\eta) \right]\approx 16\epsilon.
\end{gather}
With the BICEP2 result $r=0.2$, the energy scale of inflation is $\Lambda\sim 2\times 10^{16}$ Gev
and the slow roll parameter $\epsilon\sim 0.0125$ to the first order approximation. If $\xi^2\ll \epsilon$,
then the second order correction for the scalar spectral index $n_s$ in Eq. (\ref{nsdef}) is negligible,
we have
\begin{equation}
n'_s=\frac{r(n_s-1)}{2} + \frac{3r^2}{32} -2 \xi^2.
\end{equation}
It is clear that $n_s'\sim 10^{-3}$ with the observational results.
Therefore, to get $n_s'\sim -0.02$, we need to consider large $\xi^2$ and the second order correction
to the scalar spectral index $n_s$ in Eq. (\ref{nsdef}).
If $\xi^2\sim 0.01$, then the main contribution
to the running of the spectral comes from $\xi^2$.
For slow roll parameters $\epsilon$ and $\eta$, we have $|\epsilon| \le 0.01$ and $|\eta| \le 0.01$.
Note that $8(C+\frac{2}{3})\simeq -0.506667$, we can neglect the term $8 (C+\frac{2}{3})(2\epsilon-\eta)$
at the next leading order in Eqs. (\ref{ntdef}) and (\ref{rdef}). Thus, we will take the leading order
approximation $n_t=-2\epsilon$ and $r=16 \epsilon $ for simplicity.

The number of e-folds before the end of inflation is given by
\begin{equation}
\label{efolddef}
N(\phi)=\int_t^{t_e}Hdt\approx \frac{1}{M_{pl}^2}\int_{\phi_e}^\phi\frac{V(\phi)}{V_\phi(\phi)}d\phi=\frac{1}{\sqrt{2}\,M_{pl}}\int_{\phi_e}^\phi\frac{d\phi}{\sqrt{\epsilon}},
\end{equation}
where the value $\phi_e$ of the inflaton at the end of inflation is defined by $\epsilon(\phi_e)=1$.
Now let us briefly consider the Lyth bound \cite{Lyth:1996im}. From the above equation, we have
\begin{equation}
\Delta \phi \equiv |\phi_*-\phi_e| > {\sqrt{2\epsilon_{\rm min}}} N(\phi) M_{pl}~,~\,
\end{equation}
where $\epsilon_{\rm min}$ is the minimal $\epsilon$ during inflation.
If $\epsilon(\phi)$ is a monotonous function of $\phi$, we have
$\epsilon_{\rm min}=\epsilon(\phi_*)\equiv\epsilon$. With the BICEP2 result $r=0.16/0.20$,
we can obtain the large field inflation because of $\Delta\phi > 7 M_{pl}$.
Thus, to violate the Lyth bound and have the magnitude of $\phi$ smaller than the reduced Planck scale
during inflation, we require that $\epsilon(\phi)$ is not a monotonous function and it
has a minimum between $\phi_*$ and $\phi_e$.

\section{Single Field Inflation Models with Simple Potentials}

\subsection{Inflaton Potentials }

Herein, we will describe one type of the single field inflation models with simple potentials.
The inflation models with potential $\alpha \phi^n e^{-\beta \phi^m}$~\cite{Li:2013nfa} have been studied
systematically previously, while such type of potentials may have the
unlikeliness problem~\cite{Ijjas:2013vea} unless both $n$ and $m$ are even. Conversely,
for the S-dual inflation with the potential
$V(\phi)=V_0 {\rm sech}(\phi/M)$~\cite{Anchordoqui:2014uua}, the slow-roll parameters are
\begin{gather}
\label{sdualsl1}
\epsilon=\frac{M^2_{pl}}{2M^2}\tanh^2(\phi/M),\\
\label{sdualsl2}
\eta=4\epsilon-\frac{M^2_{pl}}{M^2},\\
\label{sdualsl3}
\xi^2=24\epsilon^2-10\left(\frac{M_{pl}}{M}\right)^2\epsilon.
\end{gather}
To satisfy slow-roll condition, $g=M/ M_{pl}$ must be large, then $\epsilon<1$ always and
inflation will not end. Thus we need another mechanism to end inflation.
For the S-dual inflation, we have
\begin{gather}
\label{sdualr}
r=16\epsilon=8g^{-2}\tanh^2(\phi/M)\le 8/g^2,\\
\label{sdualns}
n_s=1+\frac{r}{8}-\frac{2}{g^2}\le 1-\frac{r}{8},\\
\label{sdualdns}
n_s'=\frac{r}{4}\left(\frac{1}{g^2}-\frac{r}{8}\right)\ge 0.
\end{gather}
The number of e-folds before the end of inflation is given as
\begin{equation}
\label{sdulne}
N(\phi)=\int_{\phi_e}^\phi\frac{g^2}{\tanh(\phi/M)}d\phi=g^2\,\ln\left[\frac{\sinh(\phi/M)}{\sinh(\phi_e/M)}\right].
\end{equation}
If we take $\phi_e=M$, $g=M/M_{pl}=5.7735$ and $\phi/M=2.659$, we get $n_s=0.969$, $r=0.235$,
$n_s'=3.4\times 10^{-5}$ and $N=60$, which is marginally consistent with the observational
constraint at the 95\% confidence level.

\begin{figure}[htp]
\centerline{\includegraphics[width=0.4\textwidth]{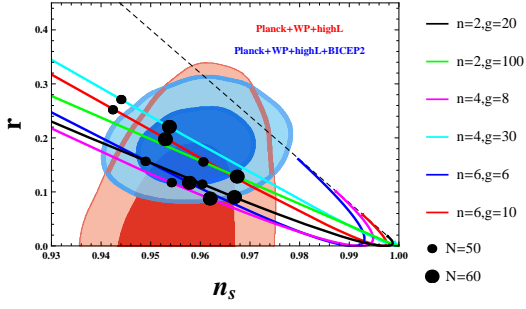}}
\caption{The $n_s-r$ diagrams for the potential $V(\phi)=V_0 \phi^n{\rm sech}(\phi/M)$.
 Confidence levels of 68\% and 95\% confidence contour from the combinations of {\em Planck}+WP+highL \cite{Ade:2013zuv,Ade:2013uln}
and {\em Planck}+WP+highL+BICEP2 data \cite{Ade:2014xna}
are also included.}
\label{sdualnsr2}
\end{figure}

\begin{figure}[htp]
\centerline{\includegraphics[width=0.4\textwidth]{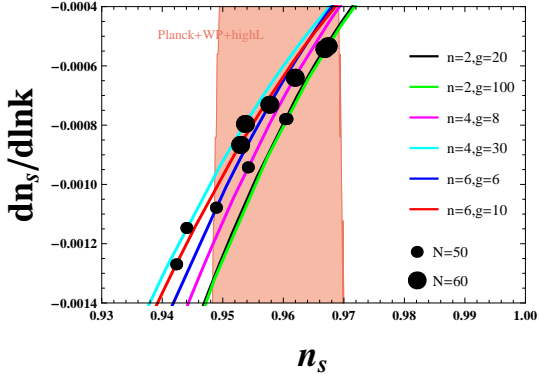}}
\caption{The $n_s-n_s'$ diagrams for the potential $V(\phi)=V_0 \phi^n{\rm sech}(\phi/M)$.
Confidence level of 95\% confidence contour from the combinations of {\em Planck}+WP+highL data \cite{Ade:2013zuv,Ade:2013uln} is also included.}
\label{sdualrun2}
\end{figure}

Let us suggest one type of the inflation models with
the hybrid monomial and S-dual potentials
 $V(\phi)=V_0 \phi^n{\rm sech}(\phi/M)$, here $n$ is an even integer. The slow-roll parameters are
\begin{gather}
\label{sdualsl4}
\epsilon=\frac{1}{2g^2}\left[\frac{n}{\phi/M}-\tanh(\phi/M)\right]^2,\\
\label{sdualsl5}
\eta=\frac{1}{g^2}\left[\frac{n(n-1)}{(\phi/M)^2}-\frac{2n}{\phi/M}\tanh(\phi/M)+\right.\nonumber\\
2\tanh^2(\phi/M)-1],\\
\label{sdualsl6}
\xi^2=\frac{1}{g^4}\left[6\tanh^4(\phi/M)-\frac{12n}{\phi/M}\tanh^3(\phi/M)+\right.\nonumber\\
\left(\frac{9n^2-3n}{(\phi/M)^2}-5\right)\tanh^2(\phi/M)+\nonumber\\
\left(\frac{8n}{\phi/M}-\frac{4n^3-6n^2+2n}{(\phi/M)^3}\right)\tanh(\phi/M)+\nonumber\\
\left.\frac{n^2(n-1)(n-2)}{(\phi/M)^4}-\frac{3n^2}{(\phi/M)^2}\right].
\end{gather}
So the spectral index also satisfies the bound
\begin{gather}
\label{sdualr1}
\eta=2\epsilon+g^{-2}\left[\tanh^2(\phi/M)-\frac{n}{(\phi/M)^2}-1\right]\le 2\epsilon,\\
\label{sdualns1}
n_s=1+2\eta-\frac{3r}{8}\le 1-\frac{r}{8}.
\end{gather}
For $n=2$, $g=100$ and $N=60$, we obtain $\phi_e/M=0.0141$, $\phi/M=0.156$, $n_s=0.967$, $r=0.128$ and $n_s'=-5.338\times10^{-4}$.
If we choose $n=4$, $g=30$ and $N=60$, we obtain $\phi_e/M=0.0941$, $\phi/M=0.715$, $n_s=0.954$, $r=0.221$ and $n_s'=-7.960\times10^{-4}$.
For $n=6$, $g=10$ and $N=60$, we get $\phi_e/M=0.4129$, $\phi/M=2.350$, $n_s=0.953$, $r=0.198$ and $n_s'=-8.667\times10^{-4}$.
The results for $n=2$, $n=4$ and $n=6$ are shown in Figs. \ref{sdualnsr2} and \ref{sdualrun2}.
We also show the observational constraints. From Figs. \ref{sdualnsr2} and \ref{sdualrun2},
it can be seen that these models are marginally consistent with the observational results
at the 95\% confidence level. For $n=2$, if there is an increase of the value of $g$ or the value of the energy scale $M$,
then both $r$ and $|n_s'|$ increase when $n_s$ is retained, then the results are almost unchanged when $g\ge 100$.
For $n=4$ and $n=6$, if there is an increase in $g$ and $n_s$ is fixed, then $r$ increases and $n_s'$ moves closer to zero.
At the 95\% confidence level, we find that $g\ge 20$ for $n=2$, $8\le g\le 30$ for $n=4$
and $6\le g\le 10$ for $n=6$.

\subsection{Supergravity Model Building}

For a given K\"ahler potential $K$ and a superpotential $W$
in the supergravity theory, we have the following scalar potential
\begin{equation}
V=e^K\left((K^{-1})^{i}_{\bar{j}}D_i W D^{\bar{j}} \overline{W}-3|W|^2 \right)~,~
\label{sgp}
\end{equation}
where $(K^{-1})^{i}_{\bar{j}}$ is the inverse of the K\"ahler metric
$K_{i}^{\bar{j}}=\partial^2 K/\partial \Phi^i\partial{\bar{\Phi}}_{\bar{j}}$, and $D_iW=W_i+K_iW$.
Also, the kinetic term for the scalar field is
\begin{equation}
{\cal L} ~=~ K_{i}^{\bar{j}} \partial_{\mu} \Phi^i \partial^{\mu} {\bar \Phi}_{\bar{j}}~.~\,
\end{equation}

Introducing two superfields
$\Phi$ and $X$, we consider the following  K\"ahler potential and superpotential
\begin{eqnarray}
K=-\frac{1}{2}(\Phi+{\bar \Phi})^2+X{\bar X}-\delta(X{\bar X})^2~,~
\label{KP-A}
\end{eqnarray}
\begin{eqnarray}
W~=~Xf(\Phi)~.~\,
\label{SP-A}
\end{eqnarray}
Therefore, the K\"ahler potential $K$ is invariant under the following shift
symmetry~\cite{Kawasaki:2000yn, Yamaguchi:2000vm,
Yamaguchi:2001pw, Kawasaki:2001as, Kallosh:2010ug, Kallosh:2010xz, Nakayama:2013jka,
Nakayama:2013txa, Takahashi:2013cxa, Li:2013nfa} is thus
\begin{eqnarray}
\Phi\rightarrow\Phi+iCM_{pl}~,~\,
\label{SSymmetry-A}
\end{eqnarray}
with $C$ a dimensionless real parameter. In general,
the K\"ahler potential $K$ is a function of $\Phi+\Phi^{\dagger}$ and independent on
the imaginary part of $\Phi$.

We can obtain the scalar potential as follows
\begin{eqnarray}
V&=& e^K\left[ |(\Phi + {\bar \Phi})Xf(\Phi)+X \frac{\partial f(\Phi)}{\partial \Phi}|^2-3|Xf(\Phi)|^2
\right. \nonumber \\ && \left.
+|({\bar X}-2\delta X{\bar X}^2)Xf(\Phi)+f(\Phi)|^2\right]~.~\,
\end{eqnarray}
Because there is no imaginary component ${\rm Im} [\Phi]$ of $\Phi$ in the K\"ahler potential
because of the shift symmetry, the potential along ${\rm Im} [\Phi]$ is considerably flat and then
${\rm Im} [\Phi]$ is a natural inflaton candidate.
From the previous studies~\cite{Kallosh:2010ug, Kallosh:2010xz, Li:2013nfa},
 the real component ${\rm Re} [\Phi]$ of $\Phi$ and $X$ can be
stabilized at the origin during inflation, {\it i.e.}, ${\rm Re} [\Phi]=0$ and $X=0$.
Therefore, with ${\rm Im} [\Phi]=\phi/{\sqrt 2}$, we obtain the inflaton potential
\begin{eqnarray}
V~=~|f(\phi/{\sqrt 2})|^2~.~\,
\end{eqnarray}

\begin{figure}[htp]
\centerline{\includegraphics[width=0.4\textwidth]{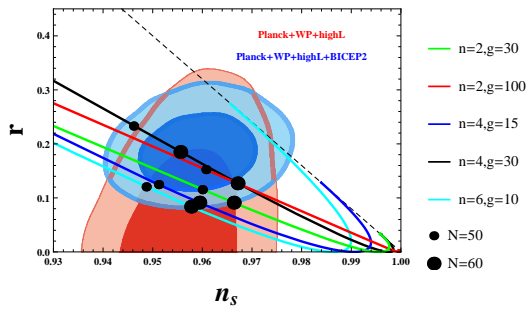}}
\caption{The $n_s-r$ diagrams for the potential $V(\phi)=V_0 \phi^n{\rm sech^2}(\phi/M)$. Confidence levels of 68\% and 95\% confidence contours from the combinations of {\em Planck}+WP+highL \cite{Ade:2013zuv,Ade:2013uln}
and {\em Planck}+WP+highL+BICEP2 data \cite{Ade:2014xna}
are also included.}
\label{hybnsr2}
\end{figure}

\begin{figure}[htp]
\centerline{\includegraphics[width=0.4\textwidth]{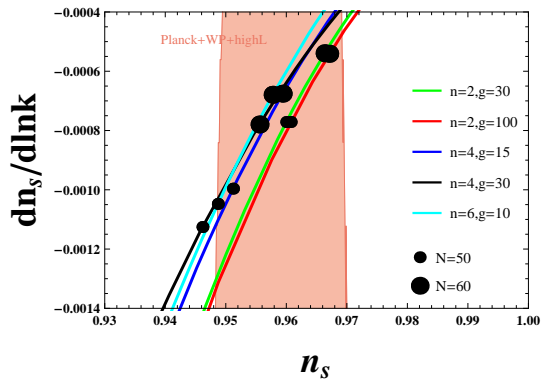}}
\caption{The $n_s-n_s'$ diagrams for the potential $V(\phi)=V_0 \phi^n{\rm sech^2}(\phi/M)$. Confidence level of 95\% confidence contour from the combinations of {\em Planck}+WP+highL data \cite{Ade:2013zuv,Ade:2013uln} is also included.}
\label{hybrun2}
\end{figure}

If we chosse $f(\Phi)$ as below
\begin{eqnarray}
f(\Phi) ~=~ \frac{{\sqrt V_0}({\sqrt 2}\Phi/M)^m}{e^{-i\Phi/{\sqrt 2}M}+e^{i\Phi/{\sqrt 2}M}}~,~\,
\end{eqnarray}
with $m$ a positive integer,
we realize the potential $V(\phi)=V_0 \phi^{n}{\rm sech^2}(\phi/M)$ with $n=2m$.
The slow-roll parameters for this type of models are
\begin{gather}
\label{sdualsl7}
\epsilon=\frac{1}{2 g^2}\left[\frac{n}{\phi/M}-2\tanh(\phi/M)\right]^2,\\
\label{sdualsl8}
\eta=\frac{1}{g^2}\left[\frac{n(n-1)}{(\phi/M)^2}-2{\rm sech^2}(\phi/M)+4\tanh^2(\phi/M)\right.\nonumber\\
\left.-\frac{4n}{\phi/M}\tanh(\phi/M)\right],\\
\label{sdualsl9}
\xi^2=\frac{1}{g^4}\left[\frac{n}{\phi/M}-2\tanh(\phi/M)\right]\times\left[\frac{n(2-3n+n^2)}{(\phi/M)^3}-\right.\nonumber\\
\frac{6n(n-1)}{(\phi/M)^2}\tanh(\phi/M)+\frac{12n}{\phi/M}\tanh^2(\phi/M)-8\tanh^3(\phi/M)+\nonumber\\
\left.2{\rm sech^2}(\phi/M)(8\tanh(\phi/M)-\frac{3n}{\phi/M})\right].
\end{gather}
Thus the spectral index also satisfies the bound
\begin{gather}
\label{sdualr2}
\eta=2\epsilon-g^{-2}\left[2{\rm sech^2}(\phi/M)+\frac{n}{(\phi/M)^2}\right]\le 2\epsilon,\\
\label{sdualns2}
n_s=1+2\eta-\frac{3r}{8}\le 1-\frac{r}{8}.
\end{gather}
For $n=2$, $g=100$, and $N=60$, we obtain $\phi_e/M=0.0141$, $\phi/M=0.155$, $n_s=0.967$, $r=0.127$ and $n_s'=-5.411\times10^{-4}$.
If we take $n=4$, $g=30$ and $N=60$, then we obtain $\phi_e/M=0.0939$, $\phi/M=0.694$, $n_s=0.956$,
$r=0.185$ and $n_s'=-7.795\times10^{-4}$.
For $n=6$, $g=10$ and $N=60$, we obtain $\phi_e/M=0.4025$, $\phi/M=2.030$, $n_s=0.958$, $r=0.084$ and $n_s'=-6.794\times10^{-4}$.
The results for $n=2$, $n=4$ and $n=6$ are shown in Figs. \ref{hybnsr2} and \ref{hybrun2}.
We also show the observational constraints. From Figs. \ref{hybnsr2} and \ref{hybrun2},
it can be seen from the models that they are also marginally consistent with the observational results at the 95\% confidence level.
For $n=2$, if $n_s$ is fixed and increase the value of $g$ or the value of the energy scale $M$,
then both $r$ and $|n_s'|$ increase, with the result almost unchanged when $g\ge 100$.
For $n=4$ and $n=6$, if we increase $g$ and retain $n_s$ fixed, then $r$ increases and $n_s'$ moves closer to zero.
At the 95\% confidence level, we find that $g\ge 30$ for $n=2$, $15\le g\le 30$ for $n=4$
and $g\sim 10$ for $n=6$.

\section{Discussion}

Herein we have proposed one type of the single field inflation models with
the hybrid monomial and S-dual potentials $V(\phi)=V_0 \phi^n{\rm sech}(\phi/M)$
and found that $n_s'$ given by the model is around $-10^{-4}$ when $n_s$ and $r$ are consistent with the
BICEP2 constraints. If we increase the model parameter $n$ or $g=M/M_{pl}$, for the
same value of $n_s$, then the tensor-to-scalar ratio $r$ increases,
but the running of the scalar spectral index $n_s'$ moves closer to zero
except for $n=2$. Therefore, the model parameters are
constrained by the observational results.
At the 95\% confidence level, we obtained that $g\ge 20$ for $n=2$, $8\le g\le 30$ for $n=4$
and $6\le g\le 10$ for $n=6$.

Then we used the supergravity model building method to propose
another type of models with the potentials $V(\phi)=V_0 \phi^n{\rm sech}^2(\phi/M)$.
The behavior of this model is similar to the inflation model
with the potential $V(\phi)=V_0 \phi^n{\rm sech}(\phi/M)$
and the model is more constrained by the observational data.
At the 95\% confidence level, we found that $g\ge 30$ for $n=2$, $15\le g\le 30$ for $n=4$
and $g\sim 10$ for $n=6$. The running of the scalar spectral index for both models is at the order
of $-10^{-4}$. Both models failed to provide the second order slow-roll parameter $\xi^2$ as large
as the first order slow-roll parameters $\epsilon$ and $\eta$. Furthermore, to obtain
large $r$, the inflaton will experience a Planck excursion because of the Lyth bound.
This is the common problem for single field inflation as suggested by Gong ~\cite{Gong:2014cqa}.
To violate the Lyth bound and result in sub-Planckian inflaton field,
the slow roll parameter $\epsilon$ needs not be retained by a monotonous function
during inflation \cite{Hotchkiss:2011gz,BenDayan:2009kv}.
Recently Ben-Dayan and Brustein ~\cite{BenDayan:2009kv} found that $n_s=0.96$, $r=0.1$ and $n_s'=-0.07$
for a single field inflation
with polynomial potential. It remains unclear if a single field inflation model can be constructed which both contain a large $r$ and $n_s'$.

\begin{acknowledgments}
This research was supported in part by the Natural Science
Foundation of China under grant numbers 11075194, 11135003, 11175270 and 11275246,
the National Basic Research Program of China (973 Program) under grant number 2010CB833000 (TL),
the Program for New Century Excellent Talents in University under grant No. NCET-12-0205
and the Fundamental Research Funds for the Central Universities under grant No. 2013YQ055.
\end{acknowledgments}


\end{document}